\documentclass[12pt]{article}
\usepackage{amssymb}

\def\hybrid{\topmargin 0pt      \oddsidemargin 0pt
        \headheight 0pt \headsep 0pt
        \voffset=-0.5cm
        \hoffset=-0.25in
        \textwidth 6.75in
        \textheight 9.5in       
        \marginparwidth 0.0in
        \parskip 5pt plus 1pt   \jot = 1.5ex}
\catcode`\@=11
\def\marginnote#1{}

\newcount\hour
\newcount\minute
\newtoks\amorpm
\hour=\time\divide\hour by60 \minute=\time{\multiply\hour by60
\global\advance\minute by-\hour}
\edef\standardtime{{\ifnum\hour<12 \global\amorpm={am}%
        \else\global\amorpm={pm}\advance\hour by-12 \fi
        \ifnum\hour=0 \hour=12 \fi
        \number\hour:\ifnum\minute<10 0\fi\number\minute\the\amorpm}}
\edef\militarytime{\number\hour:\ifnum\minute<10 0\fi\number\minute}

\def\draftlabel#1{{\@bsphack\if@filesw {\let\thepage\relax
   \xdef\@gtempa{\write\@auxout{\string
      \newlabel{#1}{{\@currentlabel}{\thepage}}}}}\@gtempa
   \if@nobreak \ifvmode\nobreak\fi\fi\fi\@esphack}
        \gdef\@eqnlabel{#1}}
\def\@eqnlabel{}
\def\@vacuum{}
\def\draftmarginnote#1{\marginpar{\raggedright\scriptsize\tt#1}}
\def\draftlabel#1{{\@bsphack\if@filesw {\let\thepage\relax
   \xdef\@gtempa{\write\@auxout{\string
      \newlabel{#1}{{\@currentlabel}{\thepage}}}}}\@gtempa
   \if@nobreak \ifvmode\nobreak\fi\fi\fi\@esphack}
        \gdef\@eqnlabel{#1}}
\def\@eqnlabel{}
\def\@vacuum{}
\def\draftmarginnote#1{\marginpar{\raggedright\scriptsize\tt#1}}

\def\draft{\oddsidemargin -.5truein
        \def\@oddfoot{\sl preliminary draft \hfil
        \rm\thepage\hfil\sl\today\quad\militarytime}
        \let\@evenfoot\@oddfoot \overfullrule 3pt
        \let\label=\draftlabel
        \let\marginnote=\draftmarginnote
   \def\@eqnnum{(\theequation)\rlap{\kern\marginparsep\tt\@eqnlabel}%
\global\let\@eqnlabel\@vacuum}  }

\def\numberbysection{\@addtoreset{equation}{section}
        \def\theequation{\thesection.\arabic{equation}}}

\def\underline#1{\relax\ifmmode\@@underline#1\else
        $\@@underline{\hbox{#1}}$\relax\fi}

\def\titlepage{\@restonecolfalse\if@twocolumn\@restonecoltrue\onecolumn
     \else \newpage \fi \thispagestyle{empty}\c@page\z@
        \def\thefootnote{\fnsymbol{footnote}} }

\def\endtitlepage{\if@restonecol\twocolumn \else  \fi
        \def\thefootnote{\arabic{footnote}}
        \setcounter{footnote}{0}}  

\numberbysection \hybrid

\newcounter{mo}

\newcommand{\ti}[1]{\tilde{#1}}

\newcommand{\vf}{\varphi}
\newcommand{\al}{\alpha}
\newcommand{\be}{\beta}

\newcommand{\om}{\omega}
\newcommand{\vth}{\vartheta}

\newcommand{\Mat}{ {\rm Mat}(N,\mathbb C) }

\newcommand{\mC}{\mathbb C}
\newcommand{\mZ}{\mathbb Z}

\newcommand{\ka}{\kappa}

\newcommand{\z}{{\zeta}}

\newcommand{\Om}{\Omega}

\newtheorem{predl}{Proposition}[section]

\def\beq{\begin{equation}}
\def\eq{\end{equation}}
\def\p{\partial}

\def\res{\mathop{\hbox{Res}}\limits}

\begin{document}

\setcounter{page}{1}


\begin{flushright}
 ITEP-TH-31/19\\
\end{flushright}
\vspace{0mm}

\begin{center}
\vspace{0mm}
 {\LARGE{Odd supersymmetrization of elliptic $R$-matrices}}
\\
\vspace{15mm} {\large \ \ {A. Levin}\,{\small $^{\S\, \ddagger}$}
 \ \ \ \ \ {M. Olshanetsky}\,{\small $^{\ddagger\, \flat\, \natural}$}
 \ \ \ \ \ {A. Zotov}\,{\small $^{\diamondsuit\, \ddagger\,
\natural}$} }
 \vspace{10mm}

 \vspace{1mm}$^\S$ - {\small{\rm National Research University Higher School of Economics, Russian Federation,   \\
 Usacheva str. 6,  Moscow, 119048, Russia}}
 \\
 \vspace{1mm} $^\ddagger$ -- {\small{\rm Institute for Theoretical and
 Experimental
 Physics of
  NRC ''Kurchatov Institute'',\\
 B. Cheremushkinskaya str. 25,  Moscow, 117218, Russia}}
 \\
 \vspace{1mm} $^\flat$ --
 {\small{\rm Institute for Information Transmission Problems RAS (Kharkevich Institute),\\
 Bolshoy Karetny per. 19, Moscow, 127994, Russia}}
\\
\vspace{1mm} $^\diamondsuit$ -- {\small{\rm
 Steklov Mathematical Institute of Russian Academy of Sciences,\\ Gubkina str. 8, Moscow,
119991,  Russia}}
 \\
  \vspace{1mm} $^\natural$ -- {\small{\rm Moscow Institute of Physics and Technology,\\ Inststitutskii per.  9, Dolgoprudny,
 Moscow region, 141700, Russia}}

\end{center}

\begin{center}\footnotesize{{\rm E-mails:}{\rm\
alevin2@hse.ru,\ olshanet@itep.ru,\ zotov@mi-ras.ru}}\end{center}
%
%

 \begin{abstract}
We study a general ansatz for an odd supersymmetric version of the
Kronecker elliptic function, which satisfies the genus one Fay
identity. The obtained result is used for construction of the odd
supersymmetric analogue for the classical and quantum elliptic
$R$-matrices. They are shown to satisfy the classical Yang-Baxter
equation and the associative Yang-Baxter equation. The quantum
Yang-Baxter is discussed as well. It acquires additional term in the
case of supersymmetric $R$-matrices.
 \end{abstract}

\bigskip
  \hfill{\em To the 80-th anniversary of Andrei Slavnov}


\section{Introduction}
\setcounter{equation}{0}

\paragraph{Kronecker function.} In this paper we deal with the Kronecker elliptic function
\cite{Weil}
 defined on the elliptic curve $\Sigma_\tau={\mC}/(\mZ\oplus\tau\mZ)$
 with the moduli $\tau$. It is fixed by the residue
  \beq\label{c06}
  \begin{array}{c}
  \displaystyle{
 \res\limits_{z=0}\phi(\hbar,z)=1
 }
 \end{array}
 \eq
and the quasi-periodic boundary conditions on the lattice
$\mZ\oplus\tau\mZ$:
  \beq\label{c07}
  \begin{array}{c}
  \displaystyle{
 \phi(\hbar,z+1)=\phi(\hbar,z)\,,\quad\quad \phi(\hbar,z+\tau)=e^{-2\pi\imath
 \hbar}\phi(\hbar,z)\,.
 }
 \end{array}
 \eq
 Explicit expression is given in terms of the Riemann
 theta-function
  \beq\label{c01}
  \begin{array}{c}
  \displaystyle{
\phi(\hbar,z;\tau)\equiv\phi(\hbar,z)=\frac{\vth'(0)\vth(\hbar+z)}{\vth(\hbar)\vth(z)}\,,
 }
 \\ \ \\
  \displaystyle{
\vth(z;\tau)\equiv\vth(z)=\displaystyle{\sum _{k\in \mathbb Z}} \exp
\left ( \pi \imath \tau (k+\frac{1}{2})^2 +2\pi \imath
(z+\frac{1}{2})(k+\frac{1}{2})\right )\,,\quad \vth(-z)=-\vth(z)\,.
 }
 \end{array}
 \eq
 The key properties of the function (\ref{c01}) are as follows:
 \begin{itemize}
\item the Kronecker function satisfies the genus one Fay trisecant
 identity \cite{Fay}:
  \beq\label{c03}
  \begin{array}{c}
  \displaystyle{
\phi(\hbar_1,z_{12})\phi(\hbar_2,z_{23})=\phi(\hbar_2,z_{13})\phi(\hbar_1-\hbar_2,z_{12})+
\phi(\hbar_2-\hbar_1,z_{23})\phi(\hbar_1,z_{13})\,,
 }
 \end{array}
 \eq
 where $z_{ij}=z_i-z_j$;
 \item the Kronecker function satisfies the heat
 equation:
  \beq\label{c10}
  \begin{array}{c}
  \displaystyle{
 2\pi\imath\p_\tau\phi(\hbar,z;\tau)=\p_z\p_\hbar\phi(\hbar,z;\tau)\,.
 }
 \end{array}
 \eq
 \end{itemize}
Using the skew-symmetry property of the Kronecker function
  \beq\label{c04}
  \begin{array}{c}
  \displaystyle{
\phi(\hbar,z_{12})=-\phi(-\hbar,z_{21})
 }
 \end{array}
 \eq
 we can rewrite (\ref{c03}) in the form
  \beq\label{c05}
  \begin{array}{c}
  \displaystyle{
\phi(\hbar_1,z_{12})\phi(\hbar_2,z_{23})+\phi(-\hbar_2,z_{31})\phi(\hbar_1-\hbar_2,z_{12})
+\phi(\hbar_2-\hbar_1,z_{23})\phi(-\hbar_1,z_{31})=0\,.
 }
 \end{array}
 \eq

\paragraph{Yang-Baxter equations.} Relations (\ref{c03}), (\ref{c10}) play a crucial role in
elliptic
 integrable systems and monodromy preserving equations since they
 underly the Lax representations with spectral parameter, classical
 and quantum $R$-matrix structures, Sklyanin algebras and the Knizhnik-Zamolodchikov-Bernard
 equations \cite{Baxter,Skl3,KrSkl,Fe}. From algebraic viewpoint the Fay
 identity (\ref{c05}) is the scalar version of the Fomin-Kirillov algebra
 \cite{FK} defined by the associative Yang-Baxter equation
  \beq\label{c11}
    \displaystyle{
  R^{\hbar_1}_{12}(z_{12})
 R^{{\hbar_2}}_{23}(z_{23})+R^{{-\hbar_2}}_{31}(z_{31})R_{12}^{{\hbar_1}-{\hbar_2}}(z_{12})+
 R^{{\hbar_2}-{\hbar_1}}_{23}(z_{23})R^{-\hbar_1}_{31}(z_{31})=0\,,
 }
  \eq
 where notations of the Quantum Inverse Scattering Method \cite{Skl3} are used,
 so that
  $R^{\hbar}_{ab}(z_{ab})$  ($R$-matrix) is a matrix valued function
  of the spectral parameter $z_a-z_b$ and the Planck constant $\hbar$. Put it  differently, $R_{ab}$
  is an operator in $\Mat^{\otimes 3}$
 acting non-trivially in the $a$-th and $b$-th tensor components. In
 particular, equation (\ref{c11}) is fulfilled by the properly normalized Baxter-Belavin
 elliptic $R$-matrix \cite{Pol}, which is then treated as a matrix
 generalization of the Kronecker function (\ref{c01}). Applications
 of (\ref{c11}) can be found in \cite{Kirillov,LOZ}.

 A skew-symmetric and unitary solution of (\ref{c11})  satisfies also the
 quantum Yang-Baxter equation:
  \beq\label{c09}
    \displaystyle{
R_{12}^\hbar(z_{12})R_{13}^\hbar(z_{13})R_{23}^\hbar(z_{23})=
R_{23}^\hbar(z_{23})R_{13}^\hbar(z_{13})R_{12}^\hbar(z_{12})\,.
 }
  \eq
 In the classical limit $\hbar\rightarrow 0$  it provides the
 classical Yang-Baxter equation  for the classical $r$-matrix:
  \beq\label{c08}
  \displaystyle{
  [r_{12}(z_{12}),r_{13}(z_{13})]+[r_{12}(z_{12}),r_{23}(z_{23})]+[r_{13}(z_{13}),r_{23}(z_{23})]=0\,.
  }
 \eq

\paragraph{Supersymmetrization.}
 Following \cite{Levin,Rabin} (see also \cite{DP}) we consider the supersymmetric
 elliptic curve, which is defined
 as a quotient of superspace $\mC^{1|1}$ (endowed with coordinates $z,\z$) by
(super)translations
  \beq\label{c33}
  \left\{
  \begin{array}{l}
 z\rightarrow z+1\,,
 \\
 \z\rightarrow \z\,,
 \end{array}
 \right.
  \qquad\qquad
  \left\{
  \begin{array}{l}
 z\rightarrow z+\tau+2\pi\imath\z\om\,,
 \\
 \z\rightarrow \z+2\pi\imath\om\,,
 \end{array}
 \right.
 \eq
where $\z$ is a superpartner to the coordinate $z$, and $\om$ is a
superpartner to the moduli of elliptic curve $\tau$.  The
supersymmetric
 elliptic curve is
equipped with the covariant derivative $D_\z=\p_\z+\z\p_z$,
$D_\z^2=\p_z$. In what follows we also use the Grassmann variables
$\mu_i$ as the superpartners to the parameters $\hbar_i$. Finally,
we have the following table of even and odd variables:
%
  \beq\label{c31}
  \begin{array}{cccc}
 \hbox{even variables:}\ &\  z_k\  &\  \tau\ &\  \hbar_i\
 \\
  \hbox{odd variables:}\ &\  \z_k\  &\ \om\  &\  \mu_i\
 \end{array}
 \eq
The variables $\z_k,\mu_i,\om$ are Grassmann, i.e.
  \beq\label{c191}
  \begin{array}{c}
  \displaystyle{
 \z_k^2=\mu_i^2=\om^2=0\,,\quad
 [\z_k,\z_l]_+=[\z_k,\mu_i]_+=[\mu_i,\mu_j]_+=[\z_k,\om]_+=[\om,\mu_i]_+=0\,.
 }
 \end{array}
 \eq

 In our recent paper \cite{LOZ0} we proposed an odd supersymmetric
 version of the Kronecker function (\ref{c01}). It is of the
 following form:
  \beq\label{c20}
  \begin{array}{c}
  \displaystyle{
 {\bf\Phi}(\hbar,z_1,z_2;\tau |\, \mu,\z_1,\z_2;\om)\equiv {\bf\Phi}^{\hbar|\,\mu}(z_1,z_2|\,\z_1,\z_2)
 =(\z_1-\z_2)\phi(\hbar,z_{12})+
 }
 \\ \ \\
  \displaystyle{
 +\om\p_1\phi(\hbar,z_{12})+2\pi\imath\z_1\z_2\om\p_\tau\phi(\hbar,z_{12})
 +\z_1\z_2\mu\p_1\phi(\hbar,z_{12})+\frac{1}{2}(\z_1+\z_2)\mu\om\p_1^2\phi(\hbar,z_{12})\,,
 }
 \end{array}
 \eq
 where $\p_1\phi(x,y)=\p_x\phi(x,y)$, $\p_2\phi(x,y)=\p_y\phi(x,y)$.
 And the truncated version is given by
  \beq\label{c25}
  \begin{array}{c}
  \displaystyle{
 {\bf\Phi}^{\hbar|\,0}(z_1,z_2|\,\z_1,\z_2)
 =(\z_1-\z_2)\phi(\hbar,z_{12})
 +\om\p_1\phi(\hbar,z_{12})+2\pi\imath\z_1\z_2\om\p_\tau\phi(\hbar,z_{12})
 \,.
 }
 \end{array}
 \eq
 The latter is (\ref{c20}) without two last terms. It was shown in
 \cite{LOZ0} that both functions (\ref{c20}), (\ref{c25}) satisfy
\begin{itemize}
\item the Fay identity (\ref{c05}) written as
  \beq\label{c24}
  \begin{array}{c}
  \displaystyle{
 {\bf\Phi}^{\hbar_1|\,\mu_1}_{12}{\bf\Phi}^{\hbar_2|\,\mu_2}_{23}
 +{\bf\Phi}^{-\hbar_2|\,-\mu_2}_{31}{\bf\Phi}^{\hbar_1-\hbar_2|\,\mu_1-\mu_2}_{12}
 +{\bf\Phi}^{\hbar_2-\hbar_1|\,\mu_2-\mu_1}_{23}{\bf\Phi}^{-\hbar_1|\,-\mu_1}_{31}=0\,,
 }
 \end{array}
 \eq
 where
 ${\bf\Phi}^{\hbar|\,\mu}_{ab}={\bf\Phi}^{\hbar|\,\mu}(z_a,z_b|\,\z_a,\z_b)$.
\item the supersymmetric version of the heat equation
  \beq\label{c30}
  \begin{array}{c}
  \displaystyle{
 \Big( \p_\om+2\pi\imath(\z_1+\z_2)\p_\tau
 \Big){\bf\Phi}^{\hbar|\,\mu}_{12}
 =\Big(\p_{\z_1}+\z_1\p_{z_1}-\frac12\mu\p_\hbar\Big)\p_\hbar
 {\bf\Phi}^{\hbar|\,\mu}_{12}\,.
 }
 \end{array}
 \eq
 For the truncated function (\ref{c25}) the last term is absent in
 the r.h.s. of (\ref{c30}).
 \end{itemize}
The identity (\ref{c24}) was used to construct the odd
supersymmetric version of the quantum Baxter-Belavin $R$-matrix in
the fundamental representation of ${\rm GL}_N$ group. The $R$-matrix
was shown to satisfy the associative Yang-Baxter equation
(\ref{c11}) written as
  \beq\label{c41}
  \begin{array}{c}
  \displaystyle{
{\bf R}_{12}^{\hbar_1|\,\mu_1}{\bf R}^{\hbar_2|\,\mu_2}_{23}
 +{\bf R}_{31}^{-\hbar_2|\,-\mu_2}{\bf R}_{12}^{\hbar_1-\hbar_2|\,\mu_1-\mu_2}
 +{\bf R}_{23}^{\hbar_2-\hbar_1|\,\mu_2-\mu_1}{\bf
 R}_{31}^{-\hbar_1|\,-\mu_1}=0\,,
 }
 \end{array}
 \eq
 where ${\bf R}_{ab}^{\hbar|\,\mu}={\bf
 R}_{ab}^{\hbar|\,\mu}(z_a,z_b|\,\z_a,\z_b)$ is defined through
 (\ref{c20}). At the same time the supersymmetric analogue of the classical
 $r$-matrix satisfies the classical (super)Yang-Baxter equation:
  \beq\label{c91}
  \begin{array}{c}
  \displaystyle{
 [{\bf r}_{12},{\bf r}_{13}]_+
 +
 [{\bf r}_{12},{\bf r}_{23}]_+
 +
[{\bf r}_{13},{\bf r}_{23}]_+=0\,,\quad
 {\bf r}_{ab}={\bf r}_{ab}(z_a,z_b|\,\z_a,\z_b)\,.
 }
 \end{array}
 \eq
The latter equation was introduced and studied in \cite{Kulish} and
\cite{Kirillov}.

\noindent{\bf Purpose of the paper.} In \cite{LOZ0} the function
(\ref{c20}) was derived from the two conditions. The first one is
the residue condition
  \beq\label{c341}
  \begin{array}{c}
  \displaystyle{
  \res\limits_{z_{1}=z_2}{\bf\Phi}^{\hbar|\,\mu}(z_1,z_2|\,\z_1,\z_2)=\z_1-\z_2\,.
 }
 \end{array}
 \eq
 The second one is the quasi-periodic boundary condition:
  \beq\label{c34}
  \begin{array}{l}
  \displaystyle{
 {\bf\Phi}^{\hbar|\,\mu}(z_1+1,z_2|\,\z_1,\z_2)
 ={\bf\Phi}^{\hbar|\,\mu}(z_1,z_2+1|\,\z_1,\z_2)={\bf\Phi}^{\hbar|\,\mu}(z_1,z_2|\,\z_1,\z_2)\,,
 }
 \end{array}
 \eq
 $$
  \displaystyle{
 {\bf\Phi}^{\hbar|\,\mu}(z_1+\tau+2\pi\imath\z_1\om,z_2|\,\z_1+2\pi\imath\om,\z_2)=
 \exp\Big(\!-2\pi\imath (\hbar+\mu\z_1+\pi\imath\mu\om)
  \Big){\bf\Phi}^{\hbar|\,\mu}(z_1,z_2|\,\z_1,\z_2)\,.
 }
$$
 $$
  \displaystyle{
 {\bf\Phi}^{\hbar|\,\mu}(z_1,z_2+\tau+2\pi\imath\z_2\om|\,\z_1,\z_2+2\pi\imath\om)=
 \exp\Big(2\pi\imath (\hbar+\mu\z_2-\pi\imath\mu\om)
  \Big){\bf\Phi}^{\hbar|\,\mu}(z_1,z_2|\,\z_1,\z_2)\,.
 }
$$
 It is similar to derivation of (\ref{c01}) from
 (\ref{c06})-(\ref{c07}).

 In this paper we consider a generalization of (\ref{c20}), where every
 term is multiplied by an arbitrary $\mC$-valued coefficient
 $A_1,...,A_5$ (the set of the coefficients is denoted by $A$):
  \beq\label{c22}
  \begin{array}{c}
  \displaystyle{
 {\bf\Phi}^{\hbar|\,\mu}(z_1,z_2|\,\z_1,\z_2|\,A)
 =A_1(\z_1-\z_2)\phi(\hbar,z_{12})+A_2\om\p_1\phi(\hbar,z_{12})
 }
 \\ \ \\
  \displaystyle{
+A_3\z_1\z_2\om\p_\tau\phi(\hbar,z_{12})
 +A_4\z_1\z_2\mu\p_1\phi(\hbar,z_{12})+\frac{A_5}{2}(\z_1+\z_2)\mu\om\p_1^2\phi(\hbar,z_{12})\,.
 }
 \end{array}
 \eq
  The
 function (\ref{c20}) is reproduced when $A_1=A_2=A_4=A_5=1$, $A_3=2\pi\imath$ and the
 truncated function (\ref{c25}) appears for $A_1=A_2=1$, $A_3=2\pi\imath$ and
 $A_4=A_5=0$.  In what follows the conditions (\ref{c341})-(\ref{c34}) are not imposed. Otherwise
 we are left with (\ref{c20}) or (\ref{c25}) only.

 The paper is organized as follows. In next Section we study
 expression (\ref{c22}) as an ansatz for the Fay identity
 (\ref{c24}) and supersymmetric version of the heat equation
 (\ref{c30}).
 Equations (\ref{c24}) and (\ref{c30}) provide conditions for the
 coefficients.  In particular, we will show that (\ref{c24}) holds true iff
 $A_1A_5=A_2A_4$. As a by product we include two $\mC$-valued
 parameters $k,\kappa$ into the heat equation (\ref{c30}):
  \beq\label{c60}
  \begin{array}{c}
  \displaystyle{
 \Big( \kappa\p_\om+2\pi\imath(\z_1+\z_2)\p_\tau
 \Big){\bf\Phi}^{\hbar|\,\mu}(A)
 =\Big(\p_{\z_1}+\z_1\p_{z_1}-\frac{k}{2}\mu\p_\hbar\Big)\p_\hbar
 {\bf\Phi}^{\hbar|\,\mu}(A)\,.
 }
 \end{array}
 \eq
  In Section
 \ref{sect3} we describe construction of elliptic $R$-matrices based
 on (\ref{c22}). This leads to additional constraint for the
 coefficients $A_k$. Besides the classical and associative
 Yang-Baxter equations (\ref{c41}), (\ref{c91}) we also discuss the
 quantum Yang-Baxter equation (\ref{c09}). It acquires additional
 term in the case of supersymmetric $R$-matrix.
A summary of results is
 given in Conclusion.

\paragraph{Acknowledgments.} The work
was supported in part by RFBR grants 18-02-01081 (A. Levin and M.
Olshanetsky), 18-01-00273 (A. Zotov) and by joint RFBR project
19-51-18006 Bolg$_a$ (M. Olshanetsky). The work of A. Levin was
partially supported by Laboratory of Mirror Symmetry NRU HSE, RF
Government grant, ag. 14.641.31.0001.
 The research of A. Zotov was also supported in part by the Young Russian Mathematics award.

\section{Generalized ansatz for the Kronecker function}\label{sect2}
\setcounter{equation}{0}

Consider the function (\ref{c22}) depending on five arbitrary
coefficients. As mentioned earlier we do not impose the
quasi-periodic boundary conditions (\ref{c34}). This is due to the
following statement, which  is verified by direct calculation:
 \begin{predl}
 The function (\ref{c22}) satisfies the boundary conditions (\ref{c34})   iff
  \beq\label{c36}
  \begin{array}{c}
  \displaystyle{
  A_1=A_2=A_4=A_5\,,\quad A_3=2\pi\imath A_1\,.
 }
 \end{array}
 \eq
 For the case when the variable $\mu$ is absent, the conditions
 (\ref{c34}) provide the truncated function (\ref{c25}).
\end{predl}
%

The purpose of the Section is to find out if the generalized
function (\ref{c22}) satisfies the Fay identity (\ref{c24}) and the
supersymmetric heat equation (\ref{c60}).
\subsection{Fay identity} The main result of the paragraph is
formulated as follows:
 \begin{predl}\label{predl1}
  The Fay identity (\ref{c24}) holds true for the function (\ref{c22}) iff
  \beq\label{c23}
  \begin{array}{c}
  \displaystyle{
  A_1A_5=A_2A_4\,.
 }
 \end{array}
 \eq
 \end{predl}
\noindent\underline{\em{Proof:}}\quad
 Verification is straightforward. Using notations for derivatives from (\ref{c20}) and
  the (anti)commutation relations (\ref{c91}) let us write down the first
 term from (\ref{c24}):
  \beq\label{b01}
  \begin{array}{c}
  \displaystyle{
  {\bf\Phi}^{\hbar_1|\,\mu_1}_{12}(z_1,z_2|\,\z_1,\z_2|\,A){\bf\Phi}^{\hbar_2|\,\mu_2}_{23}(z_2,z_3|\,\z_2,\z_3|\,A)=
 }
 \\ \ \\
   \displaystyle{
 =A_1^2(\z_1-\z_2)(\z_2-\z_3)\phi(\hbar_1,z_{12})\phi(\hbar_2,z_{23})
 +A_1A_2(\z_1-\z_2)\om\phi(\hbar_1,z_{12})\p_1\phi(\hbar_2,z_{23})+
 }
 \\ \ \\
   \displaystyle{
 +A_1A_3\z_1\z_2\z_3\om\phi(\hbar_1,z_{12})\p_\tau\phi(\hbar_2,z_{23})
 +\frac12A_1A_5(\z_1\!-\!\z_2)(\z_2\!+\!\z_3)\mu_2\om\phi(\hbar_1,z_{12})\p_1^2\phi(\hbar_2,z_{23})
 }
  \\ \ \\
   \displaystyle{
 +A_1A_4\z_1\z_2\z_3\mu_2\phi(\hbar_1,z_{12})\p_1\phi(\hbar_2,z_{23})
 +A_1A_2\om(\z_2-\z_3)\p_1\phi(\hbar_1,z_{12})\phi(\hbar_2,z_{23})+
 }
 \end{array}
 \eq
 $$
   \begin{array}{c}
  \displaystyle{
 +A_2A_4\om\z_2\z_3\mu_2\p_1\phi(\hbar_1,z_{12})\p_1\phi(\hbar_2,z_{23})
 +A_1A_3\z_1\z_2\z_3\om\p_\tau\phi(\hbar_1,z_{12})\phi(\hbar_2,z_{23})+
  }
    \\ \ \\
  \displaystyle{
 +A_1A_4\z_1\z_2\z_3\mu_1\p_1\phi(\hbar_1,z_{12})\phi(\hbar_2,z_{23})
 +\frac12A_4A_5\z_1\z_2\mu_1\z_3\mu_2\om\p_1\phi(\hbar_1,z_{12})\p_1^2\phi(\hbar_2,z_{23})+
  }
      \\ \ \\
  \displaystyle{
 +A_2A_4\z_1\z_2\mu_1\om\p_1\phi(\hbar_1,z_{12})\p_1\phi(\hbar_2,z_{23})
 +\frac12A_4A_5\z_1\z_2\z_3\mu_1\om\mu_2\p_1^2\phi(\hbar_1,z_{12})\p_1\phi(\hbar_2,z_{23})+
  }
        \\ \ \\
  \displaystyle{
 +\frac12A_1A_5(\z_1+\z_2)(\z_2-\z_3)\mu_1\om\p_1^2\phi(\hbar_1,z_{12})\phi(\hbar_2,z_{23})\,,
  }
   \end{array}
 $$
and similarly for the second and the third terms. Summing them up we
should then verify if the coefficients behind any Grassmann monomial
equals zero. For example, the coefficient behind $\z_1\z_2$,
$\z_2\z_3$ and $\z_3\z_1$ is the l.h.s. of the ordinary Fay identity
(\ref{c03}) multiplied by $A_1^2$. It is equal to zero and do not
provide any constraints for the coefficients $A_k$.

The rest of the coefficients behind Grassmann monomials vanish due
to identities obtained as some derivatives of (\ref{c03}). For
example, the coefficients behind $\z_3\om$ and $\z_1\z_2\z_3\mu_1$
vanish due to identity obtained as derivative of (\ref{c03}) with
respect to $\hbar_1$, and the coefficient behind
$\z_1\z_2\z_3\mu_1\mu_2\om$ vanishes due to the one appearing from
(\ref{c03}) by the action of
$\p_{\hbar_1}\p_{\hbar_2}(\p_{\hbar_1}+\p_{\hbar_2})$. All of them
do not impose any constrains for $A_k$ except the monomials of type
$\z^2\mu\om$. They contain the terms proportional to $A_2A_4$ and
$A_1A_5$. For each of such terms there exists an identity in the
form of some derivative of (\ref{c03}), which yields the condition
(\ref{c23}). Namely, for $\z_1\z_2\om\mu_1$ one should apply the
identity
 $(\p_{\hbar_1}^2+2\p_{\hbar_1}\p_{\hbar_2})[\hbox{(\ref{c03})}]$,
 for $\z_2\z_3\om\mu_1$ and $\z_3\z_1\om\mu_1$ --
 $\p_{\hbar_1}^2[\hbox{(\ref{c03})}]$, for $\z_1\z_2\om\mu_2$ and
 $\z_3\z_1\om\mu_2$ -- $\p_{\hbar_2}^2[\hbox{(\ref{c03})}]$.
 Finally, for the coefficient behind $\z_2\z_3\om\mu_2$ one should
 use
 $(\p_{\hbar_2}^2+2\p_{\hbar_1}\p_{\hbar_2})[\hbox{(\ref{c03})}]$.
 $\blacksquare$

\subsection{Supersymmetric heat equation} Here we prove the following
statement:
 \begin{predl}
 The heat equation (\ref{c60}) holds true for the function (\ref{c22}) iff
  \beq\label{c62}
  \begin{array}{c}
  \displaystyle{
  \kappa A_2=A_1\,,\quad \kappa A_3=2\pi\imath A_1\,,\quad
  A_4=kA_1\,, \quad \kappa A_5=kA_1\,.
 }
 \end{array}
 \eq
\end{predl}
\noindent\underline{\em{Proof:}}\quad Let us write down all terms
entering the heat equation (\ref{c60}):
  \beq\label{b11}
  \begin{array}{c}
  \displaystyle{
  \ka\p_\om{\bf\Phi}^{\hbar|\,\mu}_{12}(A)=\ka A_2\p_1\phi(\hbar,z_{12})
  +\ka A_2\z_1\z_2\p_\tau\phi(\hbar,z_{12})+\frac{\ka
  A_5}{2}(\z_1+\z_2)\mu\p_1^2\phi(\hbar,z_{12})\,,
 }
 \end{array}
 \eq
  \beq\label{b12}
  \begin{array}{c}
  \displaystyle{
  2\pi\imath(\z_1+\z_2)\p_\tau{\bf\Phi}^{\hbar|\,\mu}_{12}(A)=
  -4\pi\imath A_1\z_1\z_2\p_\tau\phi(\hbar,z_{12})
  +2\pi\imath A_2 (\z_1+\z_2)\om\p_\tau\p_1\phi(\hbar,z_{12})\,,
 }
 \end{array}
 \eq
  \beq\label{b13}
  \begin{array}{c}
  \displaystyle{
  \p_{\z_1}\p_\hbar{\bf\Phi}^{\hbar|\,\mu}_{12}(A)=
  A_1\p_1\phi(\hbar,z_{12})+A_3\z_2\om\p_\tau\p_1\phi(\hbar,z_{12})+A_4\z_2\mu\p_1^2\phi(\hbar,z_{12})+\qquad\quad
  }
  \\
  \displaystyle{
  +\frac12A_5\mu\om\p_1^3\phi(\hbar,z_{12})\,,
 }
 \end{array}
 \eq
  \beq\label{b14}
  \begin{array}{c}
  \displaystyle{
  \z_1\p_{z_1}\p_\hbar{\bf\Phi}^{\hbar|\,\mu}_{12}(A)=
 A_2\z_1\om\p_1^2\p_2\phi(\hbar,z_{12})  -A_1\z_1\z_2\p_1\p_2\phi(\hbar,z_{12})
  \!+\!\frac12A_5\z_1\z_2\mu\om\p_1^3\p_2\phi(\hbar,z_{12})\,,
 }
 \end{array}
 \eq
  \beq\label{b15}
  \begin{array}{c}
  \displaystyle{
  \mu\p_\hbar^2{\bf\Phi}^{\hbar|\,\mu}_{12}(A)=A_1\mu(\z_1-\z_2)\p_1^2\phi(\hbar,z_{12})
  +A_2\mu\om\p_1^3\phi(\hbar,z_{12})
  +A_3\z_1\z_2\mu\om\p_\tau\p_1^2\phi(\hbar,z_{12})\,.
 }
 \end{array}
 \eq
 Plugging these expressions into (\ref{c60}) we again should verify
 if the relation holds true for every Grassmann monomial including
 the trivial one. The vanishing of the coefficient behind the
 trivial monomial provides relation $A_1=\ka A_2$, and for those
 behind $\z_1\mu$ and $\z_2\mu$ -- we get the constraint $A_4=kA_1=\ka
 A_5$. The rest of the coefficients requires also to use the
 ordinary heat equation (\ref{c10}). $\blacksquare$

This result contains the previously obtained statements from
\cite{LOZ0} as particular cases.
 For example, when $k=0$ and $\kappa=1$ the last
term in the r.h.s. of equation (\ref{c60}) vanishes, and the
function (\ref{c22}) turns into (\ref{c25}). In the case $k=\ka=1$
the conditions (\ref{c62}) are solved as given in (\ref{c36}), and
we come back to the function (\ref{c20}).

Notice also that the relation (\ref{c23}) coming from the Fay
identity is valid on the constraints (\ref{c62}). So that
(\ref{c62}) is a sufficient condition for both -- the Fay identity
(\ref{c24}) and the heat equation (\ref{c22}).

\section{R-matrices and Yang-Baxter equations}\label{sect3}
\setcounter{equation}{0}

In this Section we derive the Yang-Baxter equations and find out if
they provide restrictions on possible values of the coefficients
$A_k$.

\subsection{Baxter-Belavin's $R$-matrix}\label{sect31}

Let us briefly recall the widely known construction of the elliptic
 Baxter-Belavin $R$-matrix \cite{Baxter} in the fundamental representation of ${\rm
 GL}(N,\mC)$ Lie group. We deal with a special basis in $\Mat$ known
 as the sine-algebra basis. It consists
 of $N^2$ matrices
 \beq\label{b31}
 \begin{array}{c}
  \displaystyle{
 T_a=T_{a_1 a_2}=\exp\left(\frac{\pi\imath}{N}\,a_1
 a_2\right)Q^{a_1}\Lambda^{a_2}\,,\quad
 a=(a_1,a_2)\in\mZ_N\times\mZ_N\,,
 }
 \end{array}
 \eq
 defined in terms of
 \beq\label{b32}
 \begin{array}{c}
  \displaystyle{
Q_{kl}=\delta_{kl}\exp\left(\frac{2\pi
 \imath}{N}k\right)\,,\ \ \ \Lambda_{kl}=\delta_{k-l+1=0\,{\hbox{\tiny{mod}}}
 N}\,,\quad Q^N=\Lambda^N=1_{N}\,.
 }
 \end{array}
 \eq
 The latter matrices $Q,\Lambda$ can be regarded as the finite-dimensional representation of the
 Heisenberg group
 since
  \beq\label{b33}
 \begin{array}{c}
  \displaystyle{
 \exp\left(\frac{2\pi\imath}{N}\,a_1
 a_2\right)Q^{a_1}\Lambda^{a_2}=\Lambda^{a_2}Q^{a_1}\,,\quad
 a_1,a_2\in\mZ_+\,.
 }
 \end{array}
 \eq
The product of pair of basis matrices (\ref{b31}) is easily computed
from (\ref{b33}):
  \beq\label{b34}
 \begin{array}{c}
  \displaystyle{
T_\al T_\be=\kappa_{\al,\be} T_{\al+\be}\,,\ \ \
\kappa_{\al,\be}=\exp\left(\frac{\pi \imath}{N}(\be_1
\al_2-\be_2\al_1)\right)\,,
 }
 \end{array}
 \eq
 where $\al+\be=(\al_1+\be_1,\al_2+\be_2)$.

 In accordance with the numeration of basis matrices (\ref{b31}) let
 us define the set of $N^2$ functions
  \beq\label{b35}
 \begin{array}{c}
  \displaystyle{
 \vf_a(\hbar+\Omega_a,z)=\exp(2\pi\imath\frac{a_2}{N}\,z)\,\phi(\hbar+\Omega_a,z)\,,\quad
 \Omega_a=\frac{a_1+a_2\tau}{N}\,,
 }
 \end{array}
 \eq
 where $a=(a_1,a_2)\in\mZ_N\times\mZ_N$. Then the quantum elliptic $R$-matrix is
 defined as follows:
 \beq\label{b36}
 \begin{array}{c}
  \displaystyle{
R_{12}^\hbar(z)=\sum\limits_\al T_\al\otimes T_{-\al}\,
 \vf_a(\hbar+\Omega_a,z)\,.
  }
 \end{array}
 \eq
It was constructed as solution of the quantum Yang-Baxter equation
(\ref{c09}). Later  it was also shown \cite{Pol} to satisfy the
associative Yang-Baxter equation (\ref{c11}).

{\bf Remark.} Let us remark that in the definition (\ref{b35}) the
index $a=(a_1,a_2)$ was assumed to be an element of
$\mZ_N\times\mZ_N$. Let us verify that the functions (\ref{b35}) are
invariant with respect to shifts $a_{1,2}\rightarrow a_{1,2}+N$ of
indices (discrete variables). Indeed, if $a_{1}\rightarrow a_{1}+N$
then $\Om_a\rightarrow\Om_a+1$ and the function is periodic
$\vf_a(\hbar+\Omega_a,z)=\vf_a(\hbar+\Omega_a+1,z)$ due to
(\ref{c07}). For $a_{2}\rightarrow a_{2}+N$ we have
$\Om_a\rightarrow\Om_a+\tau$ and
$\vf_a(\hbar+\Omega_a,z)\rightarrow\exp(2\pi\imath
z)\vf_a(\hbar+\Omega_a+\tau,z)=\vf_a(\hbar+\Omega_a,z)$ again due to
(\ref{c07}).

 \subsection{Supersymmetric basis functions.}

 In our previous paper \cite{LOZ0} we considered the function
 (\ref{c20}). The following three equivalent definitions for the odd supersymmetric
 analogues of the basis functions (\ref{b35}) were suggested:

 1. the first one is as follows:
  \beq\label{b40}
 \begin{array}{c}
  \displaystyle{
 {\bf\Phi}_\al^{\hbar+\Omega_\al|\,\mu}(z_1,z_2|\,z_1,\z_2)=
 \exp\Big( 2\pi\imath\frac{\al_2}{N}(z_1-z_2+\z_1\z_2)
 \Big){\bf\Phi}^{\hbar+\Omega_\al|\,\mu}(z_1,z_2|\,z_1,\z_2)=
 }
 \\ \ \\
  \displaystyle{
=\Big(1+2\pi\imath\frac{\al_2}{N}\z_1\z_2\Big){\bf\Phi}^{\hbar+\Omega_\al|\,\mu}(z_1,z_2|\,z_1,\z_2)=
 }
 \\ \ \\
  \displaystyle{
=
{\bf\Phi}^{\hbar+\Omega_\al|\,\mu}(z_1,z_2|\,z_1,\z_2)+2\pi\imath\frac{\al_2}{N}\z_1\z_2\om\p_1\phi(\hbar+\Omega_\al,z_{12})\,;
  }
 \end{array}
 \eq
 2. the second is
  \beq\label{b41}
 \begin{array}{c}
  \displaystyle{
 {\bf\Phi}_\al^{\hbar+\Omega_\al|\,\mu}(z_1,z_2|\,z_1,\z_2)=
 \exp\Big( 2\pi\imath\frac{\al_2}{N}(z_1-z_2)
 \Big){\bf\Phi}^{\hbar+\Omega_\al|\,\mu+2\pi\imath\frac{\al_2}{N}\om}(z_1,z_2|\,z_1,\z_2)\,,
 }
 \end{array}
 \eq
3. and the last one is
  \beq\label{b42}
 \begin{array}{c}
  \displaystyle{
 {\bf\Phi}_\al^{\hbar+\Omega_\al|\,\mu}(z_1,z_2|\,z_1,\z_2)=
 \exp\Big( 2\pi\imath\frac{\al_2}{N}(z_1-z_2)
 \Big)\ti{\bf\Phi}_\al^{\hbar+\Omega_\al|\,\mu}(z_1,z_2|\,z_1,\z_2)\,,
 }
 \end{array}
 \eq
with
  \beq\label{b43}
  \begin{array}{c}
  \displaystyle{
 \ti{\bf\Phi}_\al^{\hbar+\Omega_\al|\,\mu}(z_1,z_2|\,z_1,\z_2)
 =(\z_1-\z_2)\vf_\al(\hbar+\Omega_\al,z_{12})+
 \om\p_1\vf_\al(\hbar+\Omega_\al,z_{12})+
 }
 \\ \ \\
  \displaystyle{
+2\pi\imath\z_1\z_2\om\frac{d}{d\tau}\vf_\al(\hbar+\Omega_\al,z_{12})
 +\z_1\z_2\mu\p_1\vf_\al(\hbar+\Omega_\al,z_{12})+
 }
 \\ \ \\
  \displaystyle{
 +\frac{1}{2}(\z_1+\z_2)\mu\om\p_1^2\vf_\al(\hbar+\Omega_\al,z_{12})\,,
 }
 \end{array}
 \eq
 where the derivative with respect to $\tau$ in the third term of (\ref{b43})
 includes also partial derivative with respect to the first argument (it contains $\Om_\al(\tau)$), and thus
 provides the same answer as in (\ref{b41}) or (\ref{b42}). The set
 of functions were shown to satisfy the following equations (Fay
 identities):
  \beq\label{b45}
  \begin{array}{c}
  \displaystyle{
 {\bf\Phi}_{\al}^{\hbar_1+\Omega_\al|\,\mu_1}(z_1,z_2|\,\z_1,\z_2){\bf\Phi}_\be^{\hbar_2+\Omega_\be|\,\mu_2}(z_2,z_3|\,\z_2,\z_3)+
 }
  \\ \ \\
  \displaystyle{
 +{\bf\Phi}_{-\be}^{-\hbar_2-\Omega_\be|\,-\mu_2}(z_3,z_1|
 \,\z_3,\z_1){\bf\Phi}_{\al-\be}^{\hbar_1-\hbar_2+\Omega_{\al-\be}|\,\mu_1-\mu_2}(z_1,z_2|\,\z_1,\z_2)+
 }
 \\ \ \\
  \displaystyle{
 +{\bf\Phi}_{\be-\al}^{\hbar_2-\hbar_1+\Omega_{\be-\al}|\,\mu_2-\mu_1}(z_2,z_3|
 \,\z_2,\z_3){\bf\Phi}_{-\al}^{-\hbar_1-\Omega_\al|\,-\mu_1}(z_3,z_1|\,\z_3,\z_1)=0\,.
 }
 \end{array}
 \eq

 The equivalence of three above definitions holds true in the case
 (\ref{c20}), i.e. in the case $A_1=A_2=A_4=A_5, A_3=2\pi\imath
 A_1$. But the definitions are not equivalent for generic coefficients $A_k$.
 Consider the set of functions:
  \beq\label{b44}
 \begin{array}{c}
  \displaystyle{
{\bf\Phi}^{\hbar+\Omega_\al|\,\mu}_{\al}(z_1,z_2|\,\z_1,\z_2|\,A,B)=
 }
 \\ \ \\
  \displaystyle{
 ={\bf\Phi}^{\hbar+\Omega_\al|\,\mu}(z_1,z_2|\,\z_1,\z_2|\,A)
 +2\pi\imath
 B\frac{\al_2}{N}\z_1\z_2\om\p_1\vf_\al(\hbar+\Omega_\al,z_{12})\,,
 }
 \end{array}
 \eq
 where $B\in\mC$ is an arbitrary coefficient. It is easy to see that
 the above definitions (\ref{b40})-(\ref{b43}) being applied to the
 function
 ${\bf\Phi}^{\hbar+\Omega_\al|\,\mu}(z_1,z_2|\,\z_1,\z_2|\,A)$
 provide

 1. $B=A_1$,

 2. $B=A_4$,

 3. $B=A_3$

\noindent respectively. In fact, any variant is possible. Moreover,
we may keep the constant $B$ to be arbitrary. It happens due to

\begin{predl}
 The set of functions (\ref{b44}) satisfy the identities (\ref{b45}) iff
 the condition (\ref{c23}) holds true, so that the second term in
 the definition (\ref{b45}) does not provide any new constraints for
 the coefficients $A_1,...,A_5,B$.
\end{predl}
\noindent\underline{\em{Proof:}}\quad The proof is similar to the
one for Proposition \ref{predl1}. In the latter we have already
proved the statement for $B=0$ case. Due to the Grassmann monomial
$\z_1\z_2\om$ the second term from (\ref{b45}) (when $B\neq 0$)
provides new terms proportional to $A_1B$ only. They are cancelled
out with the help of derivatives of the ordinary Fay identity
(\ref{c03}) with respect to $\hbar_1$ and $\hbar_2$. $\blacksquare$

 \subsection{Associative and classical Yang-Baxter equations}

 As was shown in the previous paragraph, the functions
${\bf\Phi}^{\hbar+\Omega_\al|\,\mu}_{\al}(z_1,z_2|\,\z_1,\z_2|\,A,B)$
(\ref{b44}) with arbitrary constant $B$ satisfy the Fay identities
(\ref{b45}) without any new constraints for the coefficients.

 However there is one more restriction for the coefficients. At the
 end of Section \ref{sect31} we remarked that the
 functions $\vf_a(\hbar+\Omega_a,z)$ are invariant with respect
 the shift of discrete variables (indices)
$a_{1,2}\rightarrow a_{1,2}+N$.

It is truly important by the following reason. In the Yang-Baxter
equations we multiply the basis matrices (\ref{b31}) through the
rule (\ref{b34}). This results in the appearance of sums (or
differences) of indices in tensor components of the Yang-Baxter
equations. Finally, we use the Fay identities, which also contain
the sums (or difference) of the indices. If the basis functions were
defined for the indices in the range $0\leq a_{1,2}\leq N-1$ then
the sum or difference of two indices could be out of range.
Therefore, we need to verify if the functions (\ref{b44}) are
invariant with respect to the shifts $a_{1,2}\rightarrow a_{1,2}+N$.

Because of the property
$\vf_{a_1+N,a_2}(\hbar+1+\Omega_a,z)=\vf_a(\hbar+\Omega_a,z)$ the
shift $a_{1}\rightarrow a_{1}+N$ keeps the functions (\ref{b44})
invariant. The shift $a_{2}\rightarrow a_{2}+N$ provides non-trivial
additional terms:
  \beq\label{b46}
 \begin{array}{c}
  \displaystyle{
{\bf\Phi}^{\hbar+\Omega_{\al}+\tau|\,\mu}_{a_1,a_2+N}(z_1,z_2|\,\z_1,\z_2|\,A,B)=
 }
 \\ \ \\
  \displaystyle{
={\bf\Phi}^{\hbar+\Omega_{\al}|\,\mu}_{a_1,a_2}(z_1,z_2|\,\z_1,\z_2|\,A,B)+
2\pi\imath
 (B-A_3)\frac{a_2}{N}\z_1\z_2\om\p_1\vf_a(\hbar+\Omega_a,z_{12})\,.
 }
 \end{array}
 \eq
Finally, we conclude that in order to have invariance of the
functions (\ref{b44}) with respect to the shifts $a_{1,2}\rightarrow
a_{1,2}+N$ one should impose condition
  \beq\label{b47}
 \begin{array}{c}
  \displaystyle{
 B=A_3\,.
 }
 \end{array}
 \eq
 Then the definition (\ref{b43}) is valid for the basis functions.

The classical and associative Yang-Baxter equations are proved in
the same way as in \cite{LOZ0}. Namely, introduce the odd
supersymmetric analogue of the Baxter-Belavin's $R$-matrix
(\ref{b36}):
  \beq\label{b50}
  \begin{array}{c}
  \displaystyle{
 {\bf R}_{12}^{\hbar|\,\mu}(z_1,z_2|\,\z_1,\z_2|\,A)=\sum\limits_{\al} T_\al\otimes T_{-\al}
 {\bf\Phi}_{\al}^{\hbar+\Omega_\al|\,\mu}(z_1,z_2|\,\z_1,\z_2|\,A,B)\left.\right|_{B=A_3}\,.
 }
 \end{array}
 \eq
 This $R$-matrix satisfies the associative Yang-Baxter equation
  \beq\label{b51}
  \begin{array}{c}
  \displaystyle{
{\bf R}_{12}^{\hbar_1|\,\mu_1}{\bf R}^{\hbar_2|\,\mu_2}_{23}
 +{\bf R}_{31}^{-\hbar_2|\,-\mu_2}{\bf R}_{12}^{\hbar_1-\hbar_2|\,\mu_1-\mu_2}
 +{\bf R}_{23}^{\hbar_2-\hbar_1|\,\mu_2-\mu_1}{\bf R}_{31}^{-\hbar_1|\,-\mu_1}=0
 }
 \end{array}
 \eq
with ${\bf R}_{ab}^{\hbar|\,\mu}={\bf
 R}_{ab}^{\hbar|\,\mu}(z_a,z_b|\,\z_a,\z_b|\,A)$.

Introduce similarly the odd supersymmetric analogue of the classical
elliptic $r$-matrix
  \beq\label{b53}
  \begin{array}{c}
  \displaystyle{
{\bf r}_{12}(z_1,z_2|\,\z_1,\z_2|\,A)=\sum\limits_{\al\neq 0}
T_\al\otimes T_{-\al}
{\bf\Phi}_{\al}^{\Omega_\al|\,0}(z_1,z_2|\,\z_1,\z_2|\,A,B=A_3)\,.
 }
 \end{array}
 \eq
The function
${\bf\Phi}_{\al}^{\Omega_\al|\,0}(z_1,z_2|\,\z_1,\z_2|\,A,B=A_3)$,
where $\mu$ is replaced by 0 means that $A_4=A_5=0$. The $r$-matrix
satisfies the classical (super) Yang-Baxter equation:
  \beq\label{b54}
  \begin{array}{c}
  \displaystyle{
 [{\bf r}_{12},{\bf r}_{13}]_+
 +
 [{\bf r}_{12},{\bf r}_{23}]_+
 +
[{\bf r}_{13},{\bf r}_{23}]_+=0
 }
 \end{array}
 \eq
 with ${\bf r}_{ab}={\bf r}_{ab}(z_a,z_b|\,\z_a,\z_b|\,A)$.


\subsection{Quantum Yang-Baxter equation}


\paragraph{Non-supersymmetric case.} Let us recall how the quantum
Yang-Baxter equation (\ref{c09}) arises from the associative one
(\ref{c11}). It is enough to require the $R$-matrix to be

1. skew-symmetric, i.e.
  \beq\label{b57}
  \begin{array}{c}
  \displaystyle{
 R_{12}^{\hbar}(z)=-R_{21}^{-\hbar}(-z)
 }
 \end{array}
 \eq

2. unitary
  \beq\label{b58}
  \begin{array}{c}
  \displaystyle{
 R_{12}^{\hbar}(z)R_{21}^{\hbar}(-z)=f(h,z)1_N\otimes 1_N\,,
 }
 \end{array}
 \eq
 where $f(h,z)$ is a normalization function. For the
 Baxter-Belavin's $R$-matrix written as in (\ref{b36}) the function
 is as follows:
  \beq\label{b59}
  \begin{array}{c}
  \displaystyle{
 f(h,z)=N^2(\wp(N\hbar)-\wp(z))\,.
 }
 \end{array}
 \eq

 Indeed, consider equation (\ref{c11}) in the particular case
 $\hbar_2=\hbar_1/2$, and then make the substitution $\hbar_1\rightarrow
 2\hbar$. As a result we get
  \beq\label{b60}
    \displaystyle{
  R^{2\hbar}_{12}R^{\hbar}_{23}
  +R^{-\hbar}_{31}R_{12}^{\hbar}
  +R^{-\hbar}_{23}R^{-2\hbar}_{31}=0\,,
 }
  \eq
 where the short notations $R^{\eta}_{ab}=R^{\eta}_{ab}(z_a-z_b)$ are used.
  Multiply the latter equality by $R_{23}^\hbar$ from the left:
  \beq\label{b61}
    \displaystyle{
  R_{23}^\hbar R^{2\hbar}_{12}R^{\hbar}_{23}
  +R_{23}^\hbar R^{-\hbar}_{31}R_{12}^{\hbar}
  +R_{23}^\hbar R^{-\hbar}_{23}R^{-2\hbar}_{31}=0\,.
 }
  \eq
 By applying the skew-symmetry (\ref{b57}) to $R_{31}$ and
 the unitarity (\ref{b58}) to the expression $R_{23}^\hbar R^{-\hbar}_{23}$ in the third term we get
  \beq\label{b62}
    \displaystyle{
R_{23}^\hbar R^{\hbar}_{13}R_{12}^{\hbar}=  R_{23}^\hbar
R^{2\hbar}_{12}R^{\hbar}_{23}
  +
  f(\hbar,z_{23})R^{2\hbar}_{13}\,.
 }
  \eq
The latter equality is, in fact, particular case of more general
identities, which can be found in \cite{Pol,LOZ,LOZ2}.

 Next, consider equation (\ref{c11}) with indices 2 and 3 being
 interchanged. The latter means that we conjugate (\ref{c11}) by the
 permutation operator $P_{23}$ and redefine the variables as $z_2\leftrightarrow
 z_3$:
  \beq\label{b63}
    \displaystyle{
  R^{\hbar_1}_{13}R^{{\hbar_2}}_{32}
 +R^{{-\hbar_2}}_{21}R_{13}^{{\hbar_1}-{\hbar_2}}+
 R^{{\hbar_2}-{\hbar_1}}_{32}R^{-\hbar_1}_{21}=0\,.
 }
  \eq
Substitute again $\hbar_2=\hbar_1/2$ and denote $\hbar_1:=2\hbar$:
  \beq\label{b64}
    \displaystyle{
  R^{2\hbar}_{13}R^{\hbar}_{32}
 +R^{-\hbar}_{21}R_{13}^{\hbar}+
 R^{-\hbar}_{32}R^{-2\hbar}_{21}=0\,.
 }
  \eq
 Then, multiply the equality (\ref{b64}) by $R_{23}^\hbar$ from the
 right:
  \beq\label{b65}
    \displaystyle{
  R^{2\hbar}_{13}R^{\hbar}_{32}R_{23}^\hbar
 +R^{-\hbar}_{21}R_{13}^{\hbar}R_{23}^\hbar+
 R^{-\hbar}_{32}R^{-2\hbar}_{21}R_{23}^\hbar=0\,.
 }
  \eq
Using the skew-symmetry and unitarity we  get
  \beq\label{b66}
    \displaystyle{
 R^{\hbar}_{12}R_{13}^{\hbar}R_{23}^\hbar=
 f(\hbar,z_{23})R^{2\hbar}_{13}
 +
 R^{\hbar}_{23}R^{2\hbar}_{12}R_{23}^\hbar\,.
 }
  \eq
 Finally, the quantum Yang-Baxter equation (\ref{c09}) follows from
 comparing (\ref{b62}) and (\ref{b66}).

\paragraph{Supersymmetric case.} Let us make the calculations
similar to those from the previous paragraph for the odd
supersymmetric $R$-matrix (\ref{b50}).

First, notice that the skew-symmetry property (\ref{b57}) turns in
the supersymmetric case into the \underline{symmetry property} due
to $R$-matrix oddness:
  \beq\label{b67}
    \displaystyle{
 {\bf R}_{ab}^{\hbar|\,\mu}={\bf R}_{ba}^{-\hbar|\,-\mu}\,,
 }
  \eq
where ${\bf R}_{ab}^{\hbar|\,\mu}={\bf
 R}_{ab}^{\hbar|\,\mu}(z_a,z_b|\,\z_a,\z_b|\,A)$.

 Next, let us
 evaluate the analogue of the unitarity property (\ref{b58}):
  \beq\label{b68}
    \displaystyle{
 {\bf R}_{12}^{\hbar|\,\mu}{\bf R}_{21}^{\hbar|\,\mu}=\sum\limits_{\al,\be}
 T_\al T_{-\be}\otimes T_{-\al}T_{\be}
  {\bf\Phi}_{\al;\,12}^{\hbar+\Omega_\al|\,\mu}
  {\bf\Phi}_{\be;\,21}^{\hbar+\Omega_\be|\,\mu}
 }
  \eq
where we assume
 $
 {\bf\Phi}_{\al;\,ij}^{\hbar+\Omega_\al|\,\mu}=
 {\bf\Phi}_{\al}^{\hbar+\Omega_\al|\,\mu}(z_i,z_j|\,\z_i,\z_j|\,A,B)\left.\right|_{B=A_3}$.
 The expression in the sum can be calculated explicitly using the
 definition (\ref{b44}), (\ref{c22}). Most of the terms vanish due to (\ref{c191}).
 The non-zero terms are as follows:
  \beq\label{b69}
    \displaystyle{
 {\bf\Phi}_{\al;\,12}^{\hbar+\Omega_\al|\,\mu}
  {\bf\Phi}_{\be;\,21}^{\hbar+\Omega_\be|\,\mu}\!=
 \Big[A_1A_2(\z_1-\z_2)\om\p_\hbar+A_1A_5\z_1\z_2\mu\om\p_\hbar^2\Big]
  \vf_\al(\hbar+\Omega_\al,z_{12})\vf_\be(\hbar+\Omega_\be,z_{21})\,.
 }
  \eq
 Notice that the derivation of the latter answer used $A_1A_5=A_2A_4$ as in
 (\ref{c23}).
 Plugging (\ref{b69}) into (\ref{b68}) we see that its r.h.s. is represented
 as action of the differential operator from the quadratic brackets on
 the expression $R_{12}^\hbar(z_{12})R_{21}^\hbar(z_{21})$, which is equal to  the ordinary unitarity relation, i.e.
  \beq\label{b70}
    \displaystyle{
 {\bf R}_{12}^{\hbar|\,\mu}{\bf R}_{21}^{\hbar|\,\mu}=
 \Big[A_1A_2(\z_1-\z_2)\om\p_\hbar+A_1A_5\z_1\z_2\mu\om\p_\hbar^2\Big]R_{12}^\hbar(z_{12})R_{21}^\hbar(z_{21})\,.
 }
  \eq
 Finally, using (\ref{b59}) we get the following statement for the
  \underline{analogue of
 unitarity property}:
\begin{predl}
The analogue of the unitarity property (\ref{b58})-(\ref{b59}) for
the odd $R$-matrix (\ref{b50}) is of the form:
  \beq\label{b71}
    \displaystyle{
 {\bf R}_{12}^{\hbar|\,\mu}{\bf R}_{21}^{\hbar|\,\mu}=
 \Big(A_1A_2(\z_1-\z_2)\om N^3\wp'(N\hbar)+A_1A_5\z_1\z_2\mu\om N^4\wp''(N\hbar)\Big)1_N\otimes 1_N\,.
 }
  \eq
\end{predl}
Notice also that
  \beq\label{b72}
    \displaystyle{
 {\bf R}_{12}^{\hbar|\,\mu}{\bf R}_{21}^{\hbar|\,\mu}=-{\bf R}_{21}^{\hbar|\,\mu}{\bf
 R}_{12}^{\hbar|\,\mu}\,.
 }
  \eq

Having the properties (\ref{b67}) and (\ref{b71}) we can make the
calculations similar to the previous paragraph. The main statement
of the paragraph is as follows.
 \begin{predl}
  Consider the supersymmetric elliptic ${\rm GL}(N,\mC)$ odd $R$-matrix
  (\ref{b50}). It satisfies two equations of the Yang-Baxter type  with
  additional terms. The first one is
  \beq\label{b77}
  \begin{array}{c}
  \displaystyle{
 {\bf R}_{12}^{\hbar|\,\mu}{\bf R}_{13}^{\hbar|\,\mu}{\bf
 R}^{\hbar|\,\mu}_{23}=
 {\bf R}^{\hbar|\,\mu}_{23}{\bf R}_{13}^{\hbar|\,\mu}{\bf
 R}_{12}^{\hbar|\,\mu}+2{\bf R}^{\hbar|\,\mu}_{23}{\bf R}_{32}^{\hbar|\,\mu}{\bf
 R}_{13}^{2\hbar|\,2\mu}
 }
 \end{array}
 \eq
 or,  using it is represented as
(\ref{b71})
  \beq\label{b78}
  \begin{array}{c}
  \displaystyle{
 {\bf R}_{12}^{\hbar|\,\mu}{\bf R}_{13}^{\hbar|\,\mu}{\bf
 R}^{\hbar|\,\mu}_{23}
 =
 {\bf R}^{\hbar|\,\mu}_{23}{\bf R}_{13}^{\hbar|\,\mu}{\bf
 R}_{12}^{\hbar|\,\mu}+
 }
 \\ \ \\
  \displaystyle{
 +
  2 \Big(A_1A_2(\z_2-\z_3)\om N^3\wp'(N\hbar)+A_1A_5\z_2\z_3\mu\om N^4\wp''(N\hbar)\Big){\bf
 R}_{13}^{2\hbar|\,2\mu}
  \,.
 }
 \end{array}
 \eq
 And the second is
  \beq\label{b771}
  \begin{array}{c}
  \displaystyle{
 {\bf R}_{12}^{\hbar|\,\mu}{\bf R}_{13}^{\hbar|\,\mu}{\bf
 R}^{\hbar|\,\mu}_{23}=
 -{\bf R}^{\hbar|\,\mu}_{23}{\bf R}_{13}^{\hbar|\,\mu}{\bf
 R}_{12}^{\hbar|\,\mu} -2
  {\bf R}_{23}^{\hbar|\,\mu}{\bf R}_{12}^{2\hbar|\,2\mu}{\bf
  R}^{\hbar|\,\mu}_{23}\,.
 }
 \end{array}
 \eq
 \end{predl}
\noindent\underline{\em{Proof:}}\quad
 Consider the
associative Yang-Baxter equation (\ref{b51}) for
$\hbar_2=\hbar_1/2$, $\mu_2=\mu_1/2$, and then denote
$\hbar_1:=2\hbar$ and $\mu_1:=2\mu$:
  \beq\label{b73}
  \begin{array}{c}
  \displaystyle{
{\bf R}_{12}^{2\hbar|\,2\mu}{\bf R}^{\hbar|\,\mu}_{23}
 +{\bf R}_{31}^{-\hbar|\,-\mu}{\bf R}_{12}^{\hbar|\,\mu}
 +{\bf R}_{23}^{-\hbar|\,-\mu}{\bf R}_{31}^{-2\hbar|\,-2\mu}=0\,,
 }
 \end{array}
 \eq
 which is a direct analogue of (\ref{b61}). Multiplying it by ${\bf R}^{\hbar|\,\mu}_{23}$
 from the left and using (\ref{b67}) we obtain
  \beq\label{b74}
  \begin{array}{c}
  \displaystyle{
 {\bf R}^{\hbar|\,\mu}_{23}{\bf R}_{13}^{\hbar|\,\mu}{\bf R}_{12}^{\hbar|\,\mu}=
 -{\bf R}^{\hbar|\,\mu}_{23}{\bf R}_{12}^{2\hbar|\,2\mu}{\bf R}^{\hbar|\,\mu}_{23}
 -{\bf R}^{\hbar|\,\mu}_{23}{\bf R}_{32}^{\hbar|\,\mu}{\bf
 R}_{13}^{2\hbar|\,2\mu}\,.
 }
 \end{array}
 \eq
 Similarly to (\ref{b64}), consider the equation (\ref{b73}) with indices 2 and 3
 being interchanged
  \beq\label{b75}
  \begin{array}{c}
  \displaystyle{
{\bf R}_{13}^{2\hbar|\,2\mu}{\bf R}^{\hbar|\,\mu}_{32}
 +{\bf R}_{21}^{-\hbar|\,-\mu}{\bf R}_{13}^{\hbar|\,\mu}
 +{\bf R}_{32}^{-\hbar|\,-\mu}{\bf R}_{21}^{-2\hbar|\,-2\mu}=0\,.
 }
 \end{array}
 \eq
Multiplying it by ${\bf R}^{\hbar|\,\mu}_{23}$ from the right and
using (\ref{b67}) we obtain:
  \beq\label{b76}
  \begin{array}{c}
  \displaystyle{
 {\bf R}_{12}^{\hbar|\,\mu}{\bf R}_{13}^{\hbar|\,\mu}{\bf
 R}^{\hbar|\,\mu}_{23}=
 -
  {\bf R}_{23}^{\hbar|\,\mu}{\bf R}_{12}^{2\hbar|\,2\mu}{\bf R}^{\hbar|\,\mu}_{23}
 -
 {\bf R}_{13}^{2\hbar|\,2\mu}{\bf R}^{\hbar|\,\mu}_{32}{\bf
 R}^{\hbar|\,\mu}_{23}=
  }
 \\ \ \\
  \displaystyle{
 \stackrel{(\ref{b70}),(\ref{b72})}{=} -
  {\bf R}_{23}^{\hbar|\,\mu}{\bf R}_{12}^{2\hbar|\,2\mu}{\bf R}^{\hbar|\,\mu}_{23}
 +
 {\bf R}^{\hbar|\,\mu}_{23}{\bf
 R}^{\hbar|\,\mu}_{32}{\bf R}_{13}^{2\hbar|\,2\mu}
  \,.
 }
 \end{array}
 \eq
 {\em In contrast to the ordinary case } the r.h.s. of (\ref{b74})
 and (\ref{b76}) are not equal to each other because of the property
 (\ref{b72}). Subtracting (\ref{b76}) from (\ref{b74}) we get (\ref{b77}).

 Alternatively, we can sum up the equations (\ref{b74})
 and (\ref{b76}). Then
 the last terms in the r.h.s. are cancelled out, and we get
 (\ref{b771}). $\blacksquare$

 Let us comment on the linear $R$-matrix term, which is the last
one in the r.h.s. of (\ref{b78}). The necessity of this term becomes
obvious in the scalar ($N=1$) case. In this case we should have
${\bf \Phi}_{12}^{\hbar|\,\mu}{\bf \Phi}_{13}^{\hbar|\,\mu}{\bf
 \Phi}^{\hbar|\,\mu}_{23}
 =
- {\bf \Phi}^{\hbar|\,\mu}_{23}{\bf \Phi}_{13}^{\hbar|\,\mu}{\bf
 \Phi}_{12}^{\hbar|\,\mu}$ with the sign minus due to the odd parity of
 the permutation relating both sides. The equation with the minus
 sign is also easily follows in the scalar case from (\ref{b77})
 since $({\bf \Phi}^{\hbar|\,\mu}_{23})^2=0$. In (\ref{b78}) the
 sign behind the cubic term is plus but it is compensated with the
 linear term.

Consider also a special case of (\ref{b78}) for $A_4=A_5=0$, i.e.
when the variable $\mu$ is
 absent. Though (\ref{b78}) does not contain $A_4$ we should require $A_4=0$ since the derivation of
 (\ref{b71}) as well as the Fay identity (\ref{c23}) used the condition $A_1A_5=A_2A_4$.  Then (\ref{b78}) turns into
  \beq\label{b79}
  \begin{array}{c}
  \displaystyle{
 {\bf R}_{12}^{\hbar|\,0}{\bf R}_{13}^{\hbar|\,0}{\bf
 R}^{\hbar|\,0}_{23}
 =
 {\bf R}^{\hbar|\,0}_{23}{\bf R}_{13}^{\hbar|\,0}{\bf
 R}_{12}^{\hbar|\,0}
 +
  2 A_1A_2(\z_2-\z_3)\om N^3\wp'(N\hbar){\bf
 R}_{13}^{2\hbar|\,0}
  \,.
 }
 \end{array}
 \eq
The second term in the r.h.s. of (\ref{b79}) is proportional to
$\wp'(N\hbar)$. The function $\wp'(x)$ is double-periodic,
$\wp'(x)=-\wp'(-x)$ and has a pole of third order at $x=0$.
Therefore, it has three zeros at non-trivial half-periods
 $1/2,\tau/2,(\tau+1)/2$. At the same time the $R$-matrix ${\bf R}_{13}^{2\hbar|\,0}$
 has poles at $\hbar=\hbar_0\in\Big\{\pm\frac{1}{2N}\,,\pm\frac{\tau}{2N},\pm\frac{\tau+1}{2N}\Big\}$.
 Therefore, as a result of the substitution $\hbar=\hbar_0$ only one
 summand survives in the sum over $\al$ in (\ref{b50}). Then, according
 to (\ref{b31}), (\ref{b35}) the second term in the r.h.s. of (\ref{b79}) is proportional to
 a constant matrix
 $T_{(\pm 1,0)}\otimes 1_N\otimes T_{(\pm 1,0)}$ or $T_{(0,\pm 1)}\otimes 1_N\otimes T_{(0,\pm 1)}$
 or  $T_{(\pm 1,\pm 1)}\otimes 1_N\otimes T_{(\pm 1,\pm 1)}$
 depending on the choice of $\hbar_0$.



\section{Conclusion}
\setcounter{equation}{0}

Let us summarize the obtained results:

\begin{itemize}

\item We studied ansatz for the odd supersymmetric Kronecker
function in the form
  \beq\label{b85}
  \begin{array}{c}
  \displaystyle{
  {\bf\Phi}(\hbar,z_1,z_2;\tau |\, \mu,\z_1,\z_2;\om|\,A)\equiv
  {\bf\Phi}^{\hbar|\,\mu}_{12}=
  }
 \end{array}
 \eq
 $$
  \displaystyle{
  =\Big[A_1(\z_1-\z_2)
 +A_2\om\p_\hbar+2\pi\imath A_3\z_1\z_2\om\p_\tau
 +A_4\z_1\z_2\mu\p_\hbar+\frac{A_5}{2}(\z_1+\z_2)\mu\om\p_\hbar^2\Big]\phi(\hbar,z_1-z_2)
 }
$$
 and showed that it satisfies the Fay identity
  \beq\label{b86}
  \begin{array}{c}
  \displaystyle{
{\bf \Phi}_{12}^{\hbar_1|\,\mu_1}{\bf \Phi}^{\hbar_2|\,\mu_2}_{23}
 +{\bf \Phi}_{31}^{-\hbar_2|\,-\mu_2}{\bf \Phi}_{12}^{\hbar_1-\hbar_2|\,\mu_1-\mu_2}
 +{\bf \Phi}_{23}^{\hbar_2-\hbar_1|\,\mu_2-\mu_1}{\bf
 \Phi}_{31}^{-\hbar_1|\,-\mu_1}=0
 }
 \end{array}
 \eq
 iff $A_1A_5=A_2A_4$.

\item We considered the supersymmetric version of the heat equation
in the form
  \beq\label{b87}
  \begin{array}{c}
  \displaystyle{
 \Big( \kappa\p_\om+2\pi\imath(\z_1+\z_2)\p_\tau
 \Big){\bf\Phi}^{\hbar|\,\mu}_{12}
 =\Big(\p_{\z_1}+\z_1\p_{z_1}-\frac{k}{2}\mu\p_\hbar\Big)\p_\hbar
 {\bf\Phi}^{\hbar|\,\mu}_{12}\,.
 }
 \end{array}
 \eq
 and showed that it holds true iff the following conditions valid:
  \beq\label{b88}
  \begin{array}{c}
  \displaystyle{
  \kappa A_2=A_1\,,\quad \kappa A_3=2\pi\imath A_1\,,\quad
  A_4=kA_1\,, \quad \kappa A_5=kA_1\,.
 }
 \end{array}
 \eq

\item We defined the set of functions
  \beq\label{b89}
 \begin{array}{c}
  \displaystyle{
{\bf\Phi}^{\hbar+\Omega_\al|\,\mu}_{\al}(z_1,z_2|\,\z_1,\z_2|\,A,B)\equiv
={\bf\Phi}^{\hbar|\,\mu}_{\al;\,12}
 ={\bf\Phi}^{\hbar|\,\mu}_{12}
 +2\pi\imath
 B\frac{\al_2}{N}\z_1\z_2\om\p_1\vf_\al(\hbar+\Omega_\al,z_{12})
 }
 \end{array}
 \eq
  and proved that they satisfy the Fay identities
  \beq\label{b90}
  \begin{array}{c}
  \displaystyle{
 {\bf\Phi}_{\al;\,12}^{\hbar_1+\Omega_\al|\,\mu_1}
 {\bf\Phi}_{\be;\,23}^{\hbar_2+\Omega_\be|\,\mu_2}
 +{\bf\Phi}_{-\be;\,31}^{-\hbar_2-\Omega_\be|\,-\mu_2}
 {\bf\Phi}_{\al-\be;\,12}^{\hbar_1-\hbar_2+\Omega_{\al-\be}|\,\mu_1-\mu_2}+
 }
 \\ \ \\
  \displaystyle{
 +{\bf\Phi}_{\be-\al;\,23}^{\hbar_2-\hbar_1+\Omega_{\be-\al}|\,\mu_2-\mu_1}|
 {\bf\Phi}_{-\al;\,31}^{-\hbar_1-\Omega_\al|\,-\mu_1}=0
 }
 \end{array}
 \eq
for arbitrary constant $B$.

\item The constant $B$ is fixed to be $B=A_3$ by requirement for the
functions (\ref{b89}) to be invariant with respect to the shifts
$a_{1,2}\rightarrow a_{1,2}+N$ of indices.

\item Using (\ref{b89}), (\ref{b85}) the odd elliptic $R$-matrix
 is represented in the form:
  \beq\label{b91}
  \begin{array}{c}
  \displaystyle{
  {\bf R}_{12}(\hbar,z_1,z_2;\tau |\, \mu,\z_1,\z_2;\om|\,A)\equiv
  {\bf R}^{\hbar|\,\mu}_{12}=
 }
 \\ \ \\
  \displaystyle{
  =\Big[A_1(\z_1-\z_2)
 +A_2\om\p_\hbar+2\pi\imath A_3\z_1\z_2\om\frac{d}{d\tau}
 +A_4\z_1\z_2\mu\p_\hbar+\frac{A_5}{2}(\z_1+\z_2)\mu\om\p_\hbar^2\Big]R_{12}^\hbar\,,
}
 \end{array}
 \eq
where $R_{12}^\hbar$ is the Baxter-Belavin elliptic $R$-matrix
(\ref{b36}). It satisfies the associative Yang-Baxter equation
  \beq\label{b98}
  \begin{array}{c}
  \displaystyle{
{\bf R}_{12}^{\hbar_1|\,\mu_1}{\bf R}^{\hbar_2|\,\mu_2}_{23}
 +{\bf R}_{31}^{-\hbar_2|\,-\mu_2}{\bf R}_{12}^{\hbar_1-\hbar_2|\,\mu_1-\mu_2}
 +{\bf R}_{23}^{\hbar_2-\hbar_1|\,\mu_2-\mu_1}{\bf R}_{31}^{-\hbar_1|\,-\mu_1}=0\,.
 }
 \end{array}
 \eq
 the symmetry property (\ref{b67}) and the unitarity property
 (\ref{b71}). The odd supersymmetric version of the classical
 elliptic
 $r$-matrix (\ref{b53}) satisfies the classical (super) Yang-Baxter
 equation. Notice also that from (\ref{c06}) and (\ref{b91}) it follows that
  \beq\label{b981}
  \begin{array}{c}
  \displaystyle{
 \res\limits_{z=0}{\bf R}_{12}^{\hbar_1|\,\mu_1}=(\z_1-\z_2)A_1NP_{12}\,,
 }
 \end{array}
 \eq
 where $P_{12}$ permutation operator.

\item  The cubic relations for the supersymmetric extension of the
Baxter-Belavin's $R$-matrix have form of the quantum Yang-Baxter
equation with additional term:
  \beq\label{b95}
  \begin{array}{c}
  \displaystyle{
 {\bf R}_{12}^{\hbar|\,\mu}{\bf R}_{13}^{\hbar|\,\mu}{\bf
 R}^{\hbar|\,\mu}_{23}
 =
 {\bf R}^{\hbar|\,\mu}_{23}{\bf R}_{13}^{\hbar|\,\mu}{\bf
 R}_{12}^{\hbar|\,\mu}+
 }
 \\ \ \\
  \displaystyle{
 +
  2 \Big(A_1A_2(\z_2-\z_3)\om N^3\wp'(N\hbar)+A_1A_5\z_2\z_3\mu\om N^4\wp''(N\hbar)\Big){\bf
 R}_{13}^{2\hbar|\,2\mu}

 }
 \end{array}
 \eq
 and
  \beq\label{b96}
  \begin{array}{c}
  \displaystyle{
 {\bf R}_{12}^{\hbar|\,\mu}{\bf R}_{13}^{\hbar|\,\mu}{\bf
 R}^{\hbar|\,\mu}_{23}=
 -{\bf R}^{\hbar|\,\mu}_{23}{\bf R}_{13}^{\hbar|\,\mu}{\bf
 R}_{12}^{\hbar|\,\mu} -2
  {\bf R}_{23}^{\hbar|\,\mu}{\bf R}_{12}^{2\hbar|\,2\mu}{\bf
  R}^{\hbar|\,\mu}_{23}\,.
 }
 \end{array}
 \eq

\item Possible applications of the obtained results include
construction of the Knizhnik-Zamo\-lodchikov-Bernard (KZB) equations on
supersymmetric elliptic curves. It is the subject of our next paper \cite{LOZ20}.
 The KZ(B) equations can be treated as quantization of the monodromy preserving equations. We should mention the article
\cite{MM}, where the classical rational Schlesinger system on ${\mathbb CP}^{1|1}$ was introduced in the context of studies of the Frobenius (super)manifolds. The definition of the odd connection is similar to the rational limit of the one used in our paper.

 The obtained results allow us to describe the quantum version of the Schlesinger system (or the (q)KZ(B) equations)
 on supersymmetric elliptic curves and the corresponding deformations of the quantum Painlev\'e equations. 
 It is then also possible to evaluate the integrable deformations of the quantum (spin) Calogero-Ruijsenaars type models and the models of interacting integrable tops.
 Using the associative Yang-Baxter equation the latter integrable systems were shown to be described by the so-called $R$-matrix-valued Lax pairs, which lead to integrable long-range spin chains \cite{LOZ,SZ}. The constant coefficients studied in this paper also enter to the deformed potentials.
 The deformations coming from the Grassmann variables for these type integrable models will be studied elsewhere.
 At the same time a direct usage of the odd $R$-matrix (\ref{b91}) to the quantum inverse scattering method and construction of the spin chains with local interaction is problematic
 by two reasons. First, due to the additional terms in (\ref{b95})-(\ref{b96}), and secondly, due to the $R$-matrix 
 (\ref{b91}) is an odd operator, and therefore, it is not invertible. More promising are the deformations (via the Grassmann variables) of the ordinary elliptic $R$-matrices. We are going to study these deformations in our future publications.

\end{itemize}


\begin{small}
 
\end{small}

\end{document}